\documentclass[pdflatex,sn-mathphys-num]{sn-jnl}% Math and Physical Sciences Numbered Reference Style
\usepackage{graphicx}%
\usepackage{multirow}%
\usepackage{amsmath,amssymb,amsfonts}%
\usepackage{amsthm}%
\usepackage{mathrsfs}%
\usepackage[title]{appendix}%
\usepackage{xcolor}%
\usepackage{textcomp}%
\usepackage{manyfoot}%
\usepackage{booktabs}%
\usepackage{algorithm}%
\usepackage{algorithmicx}%
\usepackage{algpseudocode}%
\usepackage{listings}%
\theoremstyle{thmstyleone}%
\newtheorem{theorem}{Theorem}%  meant for continuous numbers
\newtheorem{proposition}{Proposition}[section]% 
\newtheorem{corollary}[theorem]{Corollary}% 

\theoremstyle{thmstyletwo}%
\newtheorem{remark}{Remark}%

\theoremstyle{thmstylethree}%
\newtheorem{definition}{Definition}%

\newcommand{\reals}{\mathbb{R}}
\newcommand{\E}{\mathbb{E}}

\usepackage{amsmath}
\usepackage{graphicx}
\usepackage{enumerate}
\usepackage{url} % not crucial - just used below for the URL 
\usepackage{longtable}

\begin{document}

\title[Spectral Clustering on Multilayer Networks with Covariates]{Spectral Clustering on Multilayer Networks with Covariates}


\author[1]{\fnm{Da} \sur{Zhao}}\email{e0546213@u.nus.edu}

\author*[1]{\fnm{Wanjie} \sur{Wang}}\email{wanjie.wang@nus.edu.sg}
\equalcont{These authors contributed equally to this work.}

\author[1]{\fnm{Jialiang} \sur{Li}}\email{jialiang@nus.edu.sg}
\equalcont{These authors contributed equally to this work.}

\affil*[1]{\orgdiv{Department of Statistics and Data Science}, \orgname{National University of Singapore}, \orgaddress{\street{S16-6-101, Science Drive 2}, \postcode{117546}, \country{Singapore}}}


\abstract{The community detection problem on multilayer networks have drawn much interest. When the nodal covariates ar also present, few work has been done to integrate information from both sources. 
To leverage the multilayer networks and the covariates, we propose two new algorithms: the spectral clustering on aggregated networks with covariates (SCANC), and the spectral clustering on aggregated Laplacian with covariates (SCALC). 
These two algorithms are easy to implement, computationally fast, and feature a data-driven approach for tuning parameter selection.

We establish theoretical guarantees for both methods under the Multilayer Stochastic Blockmodel with Covariates (MSBM-C), demonstrating their consistency in recovering community structure. Our analysis reveals that increasing the number of layers, incorporating covariate information, and enhancing network density all contribute to improved clustering accuracy. Notably, SCANC is most effective when all layers exhibit similar assortativity, whereas SCALC performs better when both assortative and disassortative layers are present.
On the simulation studies and a primary school contact data analysis, our method outperforms other methods. Our results highlight the advantages of spectral-based aggregation techniques in leveraging both network structure and nodal attributes for robust community detection.}




\maketitle

%-----------------------------------------------------------------------
\section{Introduction}\label{sec:intro}
%-----------------------------------------------------------------------

Social, economic, biological, and technological networks play a crucial role in different scientific and business studies \citep{chen2006detecting, sporns2016modular, ying2018graph, leskovec2012learning}. 
A network is usually represented by a graph $G = \{V, E\}$, where $V$ denotes a set of $N$ nodes and $E$ represents the interactions among pairs of nodes. It can be summarized by the adjacency matrix $W \in \{0, 1\}^{N \times N}$, where each entry denotes the edge existence between nodes $i$ and $j$. 
When there are several levels of interactions among the same nodes, a series of networks $G = \{V, E(l)\}_{l \in [L]}$ is observed, where $L$ is the number of layers. Correspondingly, there will be a series of adjacency matrices $\{W(l)\}_{l \in [L]}$. 
Therefore, the integration of multiple networks is of great interest \citep{de2017community, jing2021community, hu2022graph, ma2023community}. 
Besides the networks, nodal covariates also provide information, see \cite{binkiewicz2017covariate, yan2021covariate, zhang2015community, hu2024network}. Hence, a natural interest arises in the efficient and fast methods on multilayer networks with covariates. 

Our work focuses on the community detection problem. It is one of the most important problems in the social network study. Researchers are interested in recovering the community membership for all nodes, where nodes in the same community share similar connectivity patterns and properties. 
The proposed methods include Bayesian approaches \citep{nowicki2001, airoldi2008}, maximum likelihood approaches \citep{handcock2007, amini2013}, and spectral methods \citep{2012spectral, qin2013}. 
For large and relatively sparse networks, the spectral methods are of great interest due to the advantage in computational cost. 
Therefore, variants of spectral community detection methods have been explored; e.g. see \cite{joseph2016impact} and \cite{jin2015}.
Overall, these methods require the overall density of the network to be sufficiently large to achieve a consistent community detection result. 

 % This work focuses on assortative and disassortative graphs. We define an assortative graph as one in which nodes within the same cluster are more likely to share an edge than nodes in different clusters. Similarly, a disassortative graph is a type of graph where nodes tend to connect more frequently to nodes that are dissimilar to them, often from different clusters or communities.

When it comes to multilayer networks, connection information on multiple levels is collected. 
To leverage these networks, a naive aggregation approach is to take the mean of connections among all layers in \cite{han2015}. 
However, It suffers from signal cancellation when the connection probabilities in different layers are heterogeneous. To address this challenge, \cite{paul2020spectral} and \cite{bhattacharyya2018spectral} aggregate the leading eigenvectors of the adjacency matrix in each layer; \cite{lei2022biasadjusted} studies the sum-of-squared adjacency matrices with an additional bias-correction step; \cite{macdonald2022latent} introduces the common latent positions and individual latent positions for each layer, which can be estimated by convex optimization; tensor approaches are considered in \cite{lyu2022} and \cite{jing2021community}, and many other approaches. 
A related area is the dynamic networks, where both the node community membership and connection probabilities evolve. Our work focus on the multilayer networks where the community membership is the same among all layers. 

We are interested in the case that the nodal covariates are also observed. For a single-layer network, \cite{binkiewicz2017covariate} discusses the contribution of nodal covariates and network on the overall error rate. Further,  \cite{hu2024network} proves that the nodal covariates improve the community detection accuracy on the nodes with very few connections. 
When it comes to the multilayer networks, the nodal covariates are treated as static among all the layers. Under this model setting, \cite{contisciani2020} maximizes the combined likelihood of both the networks and covariates, \cite{ma2023community} investigates the information-theoretic limit achieved by the Bayes optimal estimator based on the joint signal-to-noise ratio of networks with covariates, and \cite{xu2022covariate} proposes a tensor-based community detection method by leveraging the nodal covariates to improve community detection accuracy. If we further consider the time-varying networks, \cite{guo2022time} proposes a dynamic covariate-assisted spectral clustering method, which allows covariates and community membership to change over time. 

In our work, we consider the multilayer networks with covariates where the community membership are the same on all layers. 
We propose two aggregated spectral methods: the Spectral Clustering on Aggregated Networks and Covariates (SCANC) and the Spectral Clustering on Aggregated Network Laplacian and Covariates (SCALC). 
While the aggregation methods are flexible, fast, and easy to implement, it usually suffers from the signal cancellation problem. Our SCANC algorithm works well when there are no signal cancellation across layers, and our SCALC method stands out in the signal cancellation scenario. 

We establish the theoretical consistency of both approaches. 
Considering the multilayer stochastic blockmodel with covariates, 
we present the error bounds for both approaches in Theorems \ref{thm:mis-SCL-N} -- \ref{thm:mis-SCL-N2}. The error bound is based on a combined signal-to-noise ratio of the networks and covariates among layers. We further discuss the differences between SCANC and SCALC in Theorem \ref{thm:kth}. It uncovers the veil of aggregated methods to address the signal cancellation problem. 
These results are further verified in numerical analysis.

The rest of the paper is organized as follows. In Section \ref{sec:method}, we introduce the multilayer networks with covariates data and Laplacian, then we propose the two algorithms with a discussion on tuning parameters. In Section \ref{sec:theory}, we propose the multilayer stochastic blockmodel with covariates and establish the error rate bounds of our approaches. The simulation studies can be found in Section \ref{sec:simulation}. Section \ref{sec:data} discusses the community detection results on a primary school dataset. We conclude in Section \ref{sec:disc}.

\subsection{Notations}
For an integer $L$, denote $[L]$ as $\{1, 2, \cdots, L\}$. 
For any layer-dependent quantity $q(l)$, we use $q([L])$ to denote the set $\{q(l)\}_{l \in [L]}$, and use $\tilde{q}$ to denote the average of $q(l)$ across all layers $l \in [L]$. We use $q^*(l)$ to denote the population version of $q(l)$. The exact definition varies in cases. 

For any matrix $A$, we use $\|A\|$ to denote the spectral norm and $\|A\|_F$ to denote the Frobenius norm. For any vector $x$, $\|x\|_2$ is the $\ell_2$ norm. 
For a sequence $\{a_N\}$ and $\{b_N\}$, 
$a_N = O(b_N)$ if there exists a constant $C$ so that $a_N \leq C b_N$ holds for all $N$. Denote $a_N=\Theta(b_N)$ if and only if $a_N=O(b_N)$ and $b_N=O(a_N)$. If $A\in R^{N\times N}$ is symmetric, we then denote by $\lambda_K(A)$ the $K$-th largest eigenvalue of $A$ in magnitude. Let $e_i$ denote the $i$-th standard basis vector whose $i$-th entry is 1 and the remaining entries are $0$.

%-----------------------------------------------------------------------
\section{Spectral Clustering on Aggregated Matrices}\label{sec:method}
%-----------------------------------------------------------------------

\subsection{Multilayer Networks with Covariates}\label{subsec:laplacian}

Consider the multilayer network data with covariates, denoted as $\{V, Y, W([L])\}$. 
Here, $V$ represents the node set and $Y$ represents the covariate matrix, where the $i$-th row follows the covariate vector $Y_i \in \reals^{R}$ for node $i$. 
$W(l) \in \reals^{N \times N}$ is the adjacency matrix for the $l$-th layer network, where $W_{ij}(l) = 1$ if nodes $i$ and $j$ are connected in $l$-th layer, and 0 otherwise. 
It is symmetric for an undirected network. We use $W([L])$ to denote $\{W(l)\}_{l \in [L]}$ for short. 
Suppose there is an underlying community label vector $z \in \reals^{N}$, where $z_i \in [K]$ gives the community membership of node $i$. Our goal is to recover the label vector $z$ by $Y$ and $W([L])$.  

To start, we consider the single-layer case that $L = 1$. Denote $W = W([L])$ for short. 
To recover the label vector $z$, spectral clustering is a popular method on either the covariate matrix or the adjacency matrix \citep{Lee, 2012spectral, joseph2016impact}. 
To integrate the two data sources, \cite{binkiewicz2017covariate} propose a spectral clustering method on the weighted summation of the covariate matrix and the network Laplacian. 

We first briefly introduce the covariate-assisted spectral clustering approach in \cite{binkiewicz2017covariate}. 
For the covariates, define the node-by-node similarity matrix:  
\begin{equation}\label{eqn:cov1}
    H = YY^{\top} \in \reals^{N \times N}. 
\end{equation}
For the network, let $d_i = \sum_{j = 1}^N W_{ij}$ denote the degree of node $i$ and $D \in \reals^{N \times N}$ be a diagonal matrix with the $i$-th diagonal being $d_i$. 
Given a tuning parameter $\tau$, define the regularized Laplacian matrix $\phi(W; \tau)$: 
\begin{equation}\label{eqn:lap1}
\phi(W; \tau) = (D+\tau I)^{-1/2} W (D + \tau I)^{-1/2}. 
\end{equation}
\cite{binkiewicz2017covariate} proposes the following matrix:
\begin{equation}\label{eqn:casc}
    \Phi(W, Y; \alpha, \tau) = \phi^2(W; \tau) + \alpha H. 
\end{equation}
With a proper tuning parameter $\alpha$, it's believed that $\Phi(W, Y; \alpha, \tau)$ integrates the information from both the network and covariates. Hence, a clustering approach on the leading eigenvectors of $\Phi(W, Y; \alpha, \tau)$ yields a satisfactory estimation on $z$.

These works inspire our spectral clustering approaches for the multilayer networks with covariates when $L > 1$. 
Consider the multilayer network with covariates that $\{V, W(l), Y\}_{l \in [L]}$. 
To combine the networks across layers, we propose two possible aggregated Laplacian matrices: 
\begin{eqnarray}
    \tilde{\phi}(W([L]); \tau) = \phi(\frac{1}{L}\sum_{l=1}^L W(l); \tau), \label{eqn:netLap}\\ 
   \tilde{\psi}(W([L]); \tau) = \frac{1}{L} \sum_{l=1}^L \phi^2(W(l); \tau). \label{eqn:netLap}
\end{eqnarray}
Here, $\tilde{\phi}$ is the regularized Laplacian on a simple average of all $L$ networks and $\tilde{\psi}$ is the average of all squared Laplacian matrices across all $L$ layers. When the networks across layers have the same {\it assortativity}, i.e. nodes in the same community tend to connect more/less than nodes in different communities, then $\tilde{\phi}$ works better in aggregating the signals. 
When the networks have different assortativity across layers, i.e. nodes in the same community tend to connect more than nodes in different communities in different communities on some layers, but opposite on the other layers, then $\tilde{\phi}$ suffers from the signal cancellation and $\tilde{\psi}$ works better.

The next step is to combine the networks and covariates. 
To be consistent with the network, we define $\tilde{H} = H/L$, in the form of an average.
\begin{equation}\label{eqn:cov2}
    \tilde{H} = \frac{1}{L} H = \frac{1}{L} YY^{\top}. 
\end{equation}
With preparations, we define two Laplacians for the multilayer networks with covariates:
\begin{eqnarray}
    \tilde{\Phi}(W([L]), Y; \alpha, \tau) = \tilde{\phi}^2(W([L]); \tau) + \alpha \tilde{H}, \label{eqn:lapscan}
 \\
    \tilde{\Psi}(W([L]), Y; \alpha, \tau) = \tilde{\psi}(W([L]); \tau) + \alpha \tilde{H}. \label{eqn:lapscal}
\end{eqnarray}
When $L = 1$, the two Laplacian matrices reduce to the same Laplacian matrix in \eqref{eqn:casc}. 
In the case that $L > 1$, $\tilde{\Phi}$ considers the squared Laplacian matrix on aggregated connection matrix, and $\tilde{\Psi}$ aggregates on the squared Laplacian matrices directly. By the weighted summation, the information in networks and covariates is integrated. 

Finally, we finish the introduction with some notations commonly used in subsequent analysis. 
Based on $W([L])$, the average network and corresponding degrees are defined as: 
\begin{equation}\label{eqn:avgw}
    \tilde{W} = \frac{1}{L} \sum_{l=1}^L W(l), \qquad 
    \tilde{d}_i = \sum_{j = 1}^N \tilde{W}_{ij}, \quad i \in [N]. 
\end{equation}
Let $d_i(l) = \sum_{j = 1}^N {W}_{ij}(l) $ denote the degree of node $i$ in the $l$-th layer. Then $\tilde{d}_i$ is the average of $d_i(l)$ across all $L$ layers. 

%----------------------------------------------------------------------------------------------------------------------------------------
\subsection{Spectral Clustering on Laplacian Matrices}\label{subsec:method}

The popular spectral clustering approach is computationally efficient with theoretical guarantees. It exploits information in the leading eigenvectors. Therefore, the key to successful spectral clustering is to find a proper matrix where the leading eigenvectors summarize the information well. 
We consider the newly designed Laplacian matrices in Section \ref{subsec:laplacian}. 
We first introduce the generic spectral clustering algorithm as Algorithm \ref{alg:sc}, which returns the estimated labels based on any  symmetric matrix input $A$. 

\begin{algorithm}
\caption{General Spectral Clustering Algorithm}\label{alg:sc}
\begin{algorithmic}[1]
\Require the matrix $A \in \reals^{N \times N}$; the number of communities $K$.
\Ensure the estimated label vector $\hat{z} \in [K]^N$
\State $\xi_1,\cdots, \xi_K = EigenDecompostion(A)$, corresponding to the largest
 $K$ eigenvalues in magnitude
\State $\Xi =[\xi_1, \cdots, \xi_K]\in\mathbb{R}^{N \times K}$
\State $\hat{z} = kmeans(\Xi, K)$
\end{algorithmic}
\end{algorithm}

We consider two possible choices of the input matrix $A$ for Algorithm \ref{alg:sc}. 
When $A = \tilde{\Phi}({W}([L]), Y; \alpha, \tau)$ in \eqref{eqn:lapscan}, the Laplacian on aggregated adjacency matrix and covariates, we call the algorithm as {\it the spectral clustering on aggregated networks and covariates (SCANC)}. When $A = \tilde{\Psi}(W([L]), Y; \alpha, \tau)$ in \eqref{eqn:lapscal}, the aggregated Laplacian matrix and covariates, we call the approach as {\it the spectral clustering on aggregated Laplacian and covariates (SCALC)}. As we discussed in Section \ref{subsec:laplacian}, SCANC works better when the networks enjoy the same assortativity and SCALC works better otherwise. 
We want to remind the readers that, when $L = 1$, both SCANC and SCALC degenerate as the covariate-assisted spectral clustering approach in \cite{binkiewicz2017covariate}.

Other choices of $A$ yield more spectral clustering approaches. For example, letting $A = \tilde{\phi}^2$, it becomes the spectral clustering approach on aggregated networks without covariates. 
This approach becomes the spectral clustering on network approach in \cite{joseph2016impact} when $L = 1$. 
On the other hand, $A = \tilde{H}$ results in the classical spectral clustering approach on covariates in \cite{Lee, jin2017phase}. 
A comparison of these methods explains the role of networks and covariates in community detection, which can be found in Section  \ref{sec:theory}. 

\begin{remark}
There are two tuning parameters required for SCANC and SCALC in Algorithm \ref{alg:sc}, the Laplacian regularization parameter $\tau$ and the weight $\alpha$. 
The parameter $\tau$ should be at the order of $\log N$. In numerical analysis, we set $\tau = N^{-1}\sum_{i=1}^N\tilde{d}_i$, the average aggregated degree across all nodes. The selection of $\alpha$ is discussed in Section \ref{subsec:alpha}. 
In Section \ref{sec:theory}, we prove the consistency holds for $\tau$ and $\alpha$ in the range. 
\end{remark}

%---------------------------------------------------------------
\subsection{Data-Driven Selection of the Tuning Parameter}
\label{subsec:alpha}
To calculate the aggregated Laplacian matrix, we need a tuning parameter $\alpha$ in \eqref{eqn:lapscan} and \eqref{eqn:lapscal} to balance the effects of network and covariates. 
A proper selection of $\alpha$ is required to achieve consistent clustering results. 
We start with the SCANC approach to discuss the selection of $\alpha$, and the analysis can be similarly applied to other approaches. Rigorous theoretical proof can be found in Section \ref{sec:theory} and supplementary materials \cite{supp}. 

For SCANC, the spectral clustering approach is applied to the regularized Laplacian on aggregated networks 
$\tilde{\Phi}(W([L]), Y; \alpha,\tau) = \tilde{\phi}^2(W([L]); \tau) + \alpha \tilde{H}$. 
For notation simplicity, we write $\tilde{\phi} = \tilde{\phi}(W([L]); \tau)$ for short. 
According to standard matrix perturbation theory, for both $\tilde{\phi}^2$ and $\tilde{H}$ to contribute in the leading $K$ eigenvectors of $\tilde{\Phi}(W([L]), Y; \alpha,\tau)$, $\alpha$ must be in a range so that $\lambda_1(\tilde{\phi}_1^2) > \alpha \bigl(\lambda_K(\tilde{H}) - \lambda_{K+1}(\tilde{H})\bigr)$ and $\lambda_K(\tilde{\phi}^2) - \lambda_{K+1}(\tilde{\phi}^2) < \alpha \lambda_1(\tilde{H})$ both hold. 
It naturally gives us the range $\alpha \in [\alpha_{\min}, \alpha_{\max}]$ to consider, where 
\begin{equation}\label{eqn:alpharange}
    \alpha_{\min }=\frac{\lambda_{K}(\tilde{\phi}^2)-\lambda_{K+1}(\tilde{\phi}^2)}{\lambda_{1}(\tilde{H})}, 
    \quad
\alpha_{\max}=\frac{\lambda_1(\tilde{\phi}^2)}{ \lambda_R (  \tilde{H} )1_{R\leq K}+[\lambda_K( \tilde{H} )-\lambda_{K+1}( \tilde{H} )]1_{R> K}}.
\end{equation}
Here, in the formula of $\alpha_{\max}$, we consider the case $R < K$ and $R \geq K$. 
When $R < K$, there are fewer covariates than the number of communities, the inequality is revised to $\lambda_1(\tilde{\phi}^2) > \alpha \lambda_R(\tilde{H})$. Hence $\alpha_{\max}$ is revised correspondingly. 
Theoretical discussions on the range $[\alpha_{\min}, \alpha_{\max}]$ can be found in Corollary \ref{corollary} and the remarks below. 
Both $\alpha_{\min}$ and $\alpha_{\max}$ can be calculated based on the observed data, which guarantees a good result of SCANC. 

Once the range is fixed, the exact selection of $\alpha$ is not crucial to the clustering consistency. In the algorithm, the selection is made by minimizing the loss function of the $k$-means algorithm, i.e., the within-cluster sum of squares when each row of the spectral matrix $\Xi$ is treated as a data point. 

When it comes to other approaches involving $\alpha$, e.g. SCALC, the same selection steps can be applied with corresponding matrices used in \eqref{eqn:alpharange}. 
In SCALC, $A = \tilde{\psi} + \alpha \tilde{H}$ is of interest. Therefore, $\alpha$ is selected from $[\alpha_{\min}, \alpha_{\max}]$ where $\alpha_{\min}$ and $\alpha_{\max}$ are calculated by replacing $\tilde{\phi}^2$ by $\tilde{\psi}$ in \eqref{eqn:alpharange}: 
\[
    \alpha_{\min }=\frac{\lambda_{K}(\tilde{\psi})-\lambda_{K+1}(\tilde{\psi})}{\lambda_{1}(\tilde{H})}, 
    \quad
\alpha_{\max}=\frac{\lambda_1(\tilde{\psi})}{ \lambda_R (  \tilde{H} )1_{R\leq K}+\{\lambda_K( \tilde{H} )-\lambda_{K+1}( \tilde{H} )\}1_{R> K}}.
\]

%-------------------------------------------------------------------
\section{Theoretical Guarantee}\label{sec:theory}
%-------------------------------------------------------------------

\subsection{Multilayer Stochastic Blockmodel with Covariates}\label{subsec:model}
The stochastic blockmodel (SBM) was introduced in \cite{holland1983}, and then widely used to study the community detection problem \citep{Lei2015, nowicki2001, hu2024network}. The model is then generalized to multilayer networks with fruitful results \citep{ bhattacharyya2018spectral, paul2020spectral, chen2022global, lei2022biasadjusted}. We follow this line to set up the multilayer stochastic blockmodel with covariates. 

\begin{definition}[Multilayer Stochastic Blockmodel with Covariates] 
Consider a multilayer network with covariates $\{W(l), Y\}_{l \in [L]}$. Suppose there is a community label vector $z$. 

Suppose there is a series of matrices $B(l) \in \reals^{K \times K}$, so that the adjacency matrix $W(l)$ conditional on $z$ has independent entries $W_{ij}(l) \sim Bernoulli(B_{z_i, z_j}(l))$, $i, j \in [N]$, $l \in [L]$. 
Suppose the covariate matrix $Y \in [-J, J]^{N \times R}$ has independent entries for a constant $J$, conditional on $z$. The mean covariate vector is $\E[Y_i|z_i] = M_{z_i} \in \reals^{R}$. 

We call such a model a {\it Multilayer Stochastic Blockmodel with Covariates (MSBM-C)}, with parameters $\{B([L]), R, M, J\}$. 
\end{definition}
Under this model, the community label vector $z$ is the same across all layers, which is a standard setting in multilayer network models \citep{lei2020, han2015, lei2022biasadjusted, arroyo2021inference, xu2022covariate}. Denote $Z \in \reals^{N \times K}$ as the community membership matrix, where $Z_{ik} = 1$ if $z_i = k$ and $Z_{ik} = 0$ otherwise. In the matrix form, the MSBM-C follows that 
\begin{equation}\label{eqn:sbm}
\E[W(l)] = Z B(l) Z^{\top}, \quad l \in [L], 
\quad \mbox{and } \quad
\E[Y] = Z M.
\end{equation}
Here, the mean covariate matrix $M = [M_1, \cdots, M_K]^{\top} \in \reals^{K \times R}$ has rows as mean covariate vectors of each community. 

In Section \ref{subsec:laplacian}, we have introduced the assortative and disassortative networks, which may bring signal cancellation issue. 
We define the assortative and disassortative networks here. 
\begin{definition}\label{def:ass}
    Consider a single-layer stochastic blockmodel with community by community connection matrix $B \in \reals^{K \times K}$. We call it as 
    \begin{itemize}
        \item an assortative network, if $\min_{k} B_{kk} > \max_{k_1 \neq k_2} B_{k_1, k_2}$; or 
        \item a disassorative network, if $\max_{k} B_{kk} < \min_{k_1 \neq k_2} B_{k_1, k_2}$.
    \end{itemize}
\end{definition}
In other words, for the assortative networks, the connections within a community will be denser than those across communities; and for the disassortative networks, it works in an opposite way. For a multilayer network, if the networks on some layers are assortative and on the other layers are disassorative, then a simple average of all networks will result in a signal cancellation. 

To facilitate the analysis of our new algorithms, we define the population version of $\Phi$ and $\Psi$ under the MSBM-C model. 
We first focus on the SCANC approach with $\tilde{\Phi}(W([L]), Y; \alpha, \tau)$ in \eqref{eqn:lapscan}. 
Define the average $\tilde{B} = L^{-1}\sum_{l = 1}^L B(l)$. The expected average adjacency matrix and corresponding Laplacian follow
\begin{equation}\label{eqn:oracaladj}
    \tilde{W}^* = \E[\tilde{W}] = Z^\top \tilde{B} Z, \quad 
    \tilde{\phi}^* = \phi(\tilde{W}^*; \tau).
\end{equation}
Let $\tilde{d}_i = \sum_{j=1}^N \tilde{W}_{ij}$ be the average degree and $\tilde{d}_i^* = \E[\tilde{d}_i]$. Denote $\tilde{D}^*$ the diagonal matrix with diagonals as $\tilde{d}_i^*$. Then the population version of Laplacian $\tilde{\Phi}(W([L]), Y; \alpha, \tau)$ in \eqref{eqn:lapscan} follows  
\begin{eqnarray}
\label{eqn:tml}
\tilde{\Phi}^*(\alpha, \tau) & = & (\tilde{\phi}^*)^2 + \alpha \E[\tilde{H}] \nonumber\\
& = & (\tilde{D}^* +\tau I)^{-1/2} Z \tilde{B} Z^{\top} (\tilde{D}^* + \tau I)^{-1} Z \tilde{B} Z^{\top} (\tilde{D}^* +\tau I)^{-1/2} + \alpha \E[\tilde{H}].
\end{eqnarray}

For SCALC, we need to derive the oracle Laplacian on each layer. Denote $W^*(l) = \E[W(l)] = Z^\top B(l) Z$.
Let $D(l)$ be the degree matrix for the $l$-th layer and $D^*(l) = \E[D(l)]$. It follows 
\begin{eqnarray}
%\label{eqn:oracle scal}
    \tilde{\psi}^* & = & \frac{1}{L} \sum_{l=1}^L \phi^2(W^*(l); \tau)\nonumber\\
    & = &
    \frac{1}{L} \sum_{l=1}^L (D^*(l) + \tau I)^{-\frac{1}{2}}Z B(l) Z^\top (D^*(l) + \tau I)^{-1} Z B(l) Z^\top (D^*(l) + \tau I)^{-\frac12},
\end{eqnarray}
and the Laplacian with covariates follows
\begin{equation}\label{eqn:oracle scalc}
    \tilde{\Psi}^*(\alpha, \tau) = \tilde{\psi}^* + \alpha \E[\tilde{H}].
\end{equation}

%\begin{comment}
%Based on the setting of above model, all nodes in the same community share the same expected degree. Similar to the degree matrix $\mD$ with respect to each node, we define one kind of community-wise degree matrix $\mD_B $ with respect to each community. Define $\mD_B \in \mathbb{R}^{K \times K}$ to be a diagonal matrix with diagonals as $B^* Z^{\top} \mathbf{1}_N$. Next, we define the counterpart of $\tmL(\alpha) \in \mathbb{R}^{K \times K}$, that  
%\begin{equation}
%\tmL_B(\alpha) =  (\mD_B + \tau I)^{-1/2} B^* Z^{\top} (\mD + \tau I)^{-1} Z B^* (\mD_B + \tau I)^{-1/2} + \alpha \sum_{t = 1}^T M(t) M^{\top}(t)/T^{3/2}. 
%\end{equation}
%
%Based on these notations, we introduce the Lemma\ref{lemma2.1} in appendix.
%
%The requirement on $\varkappa$ is to control $E[Y(t)Y^T(t)]-E[Y(t)] E[Y^T(t)]$, which is actually equal to diagonal matrix with $i$th diagonal as $\sum_{t=1}^T\sum_{r=1}^RVar(Y_{ir}(t))$. The noise from it should be small enough so that the information contained in $\tmL_B(\alpha)$ can be well-presented. 
%
%The lemma indicates that the eigenvector of $\tmL(\alpha)$ delivers the community information. If and only if nodes are in the same community ($Z_i = Z_j$), they will have identical rows in $\mathcal{U}$. Clustering the rows of these eigenvectors directly will provide a satisfactory community detection result. 
%
%We also consider a special case when covariates are not available and obtain the Lemma \ref{lemma2.2} about the eigen-space of $\mL_{\tau}$ under dynamic SBM as in appendix.
%\end{comment}

\subsection{SCANC: Consistency and Tuning Parameters}\label{subsec:consistency}
%
%--------------------------------------------------------------------
 %The proof of consistency for SCANC under multilayer stochastic block model contains three results. 
 %The Lemma 1.2 in the appendix shows the eigendecomposition of {\color{blue}$\tilde{\Phi}^*(\alpha, \tau)$}. 
 To discuss the consistency, we first bound the difference between the sample Laplacian $\tilde{\Phi}(W([L]), Y; \alpha, \tau)$ and the population Laplacian $\tilde{\Phi}^*(\alpha, \tau)$ in Proposition \ref{SCL-N-Laplacian}. 
 Based on this bound, we then demonstrate the difference between the sample and population eigenvectors in 
 Proposition \ref{SCL-N-eigenvector}. These results allow us to bound the mis-clustering rate of SCANC in Theorem \ref{thm:mis-SCL-N}. In 
Corollary \ref{corollary}, we elaborate the upper bound under a simplified MSBM-C model, which explains the consistency.

%------------------------------------------------------------
\begin{proposition}
\label{SCL-N-Laplacian}
Define the minimal expected degree $d=\min_i \tilde{d}^*_i$ and the overall covariate fourth moment $\tilde{w}=\frac{8}{L}\sum_{r=1}^{R} \{\sum_{i=1}^N \E[Y_{i r}^2] \sum_{j=1}^N Var(Y_{j r}) +\E[Y_{i r}^4]\}$.
 For any $\epsilon >0$, if $(d+\tau) L>3\log(8N/\epsilon)$ and $\tilde{w}>27N^2J^4\log(8N/\epsilon)/L$, then with probability at least $1-\epsilon$, 
\begin{equation}
    ||\tilde{\Phi}(W([L]), Y; \alpha, \tau)-\tilde{\Phi}^*(\alpha, \tau)||\leq \frac{12\sqrt{3/(d+\tau)}+\sqrt{3\tilde{w}\alpha^2}}{\sqrt{L}}\sqrt{\log(\frac{8N}{\epsilon})}
\end{equation}
\end{proposition}

For the sample Laplacian matrix to be close to the population Laplacian matrix, we need the network to have an overall density $dL > 3\log(8N/\epsilon)$. In other words, the sparsity condition is relaxed on each single layer. Such a phenomenon is also found in other multilayer networks literature, such as \cite{lei2020}. 

When all the covariates have constant variance, the condition holds when the number of covariates $R > \Theta(\log(N))$, increasing with the number of nodes. 
A specific example is that there is a covariate matrix $Y(l)$ on each layer $l \in [L]$, and we create a big covariate matrix $Y = [Y(1), Y(2), \cdots Y(L)]$. In this case, the number of covariates naturally increases with the number of layer $L$. Hence, increasing the number of layers will gain more information on both networks and covariates. It can be found in our numerical analysis sections.

% {\color{blue}If we consider a node-contextualized multilayer stochastic block model with two multilayer-varying blocks, within block multilayer-varying probability $p(l)$, and between block multilayer-varying probability $q(l)$. It is easy to see that the condition $(d+\tau)L>3\log(8N/\epsilon)$ holds when $\sum_{l=1}^L\{p(l)+q(l)\}>\Theta \{\log (N)/N\}$, which is a condition restrict the sparsity of the multilayer-varying graphs, and larger $L$ relax the restriction. The condition $\tilde{w}>27N^2J^4\log(8N/\epsilon)/L$ holds when $R> \Theta(\log (N))$, which is a condition that requires the number of covariates to grow with the number of nodes. %Sometimes, the covariate matrix $Y$ comprises the covariate matrices from different layers. In such cases, a larger $L$ also helps relax the restriction. 
% Next in Proposition \ref{SCL-N-eigenvector}, we bound the difference between the sample and population eigenvectors.}

%-------------------------------------------------------------
Now we define the Laplacian matrix based on the communities instead of the nodes to facilitate the analysis. Define a diagonal matrix $\tilde{D}^*_C \in \reals^{K \times K}$, where the $k$-th diagonal is $\sum_{k_1 = 1}^K n_{k_1}\tilde{B}_{k, k_1}$, the expected degree of one node in this community. Define the Laplacian matrix in this case:
\[
\tilde{\Phi}_C^*(\alpha)=(\tilde{D}_{C}^{*}+\tau I)^{-1/2} \tilde{B}Z^{\top}(\tilde{D}^{*}+\tau I)^{-1}Z\tilde{B}( \tilde{D}_{C}^{*}+\tau I)^{-1/2}+\alpha MM^{\top}/L. 
\]
For the covariates, define $c_{k}=\sum_{r=1}^{R} \operatorname{Var}(Y_{i r}|z_i = k)$ and $\bar{c}=\frac{1}{K}\sum_{k=1}^K c_{k}$.  
Denote $\varkappa = \max_{1 \leq k \leq K}|c_{k}-\bar{c}|$ be the maximum distance to the center. It evaluates the noise level in the covariates. 
Lastly, recall for any matrix $A$, $\lambda_k(A)$ denotes the $k$-th largest eigenvalue of $A$ in magnitude. 
\begin{proposition}
\label{SCL-N-eigenvector}
Let the columns of $U_{\Phi}$ and $U_{\Phi}^*$ contain the top $K$ eigenvectors of $\tilde{\Phi}(W([L]), Y; \alpha, \tau)$ and $\tilde{\Phi}^*(\alpha, \tau)$ respectively. 
Suppose all the assumptions in Proposition \ref{SCL-N-Laplacian} hold for a constant $\epsilon > 0$. Further, suppose
$\lambda_K(\tilde{\Phi}^*(\alpha, \tau)) \geq 2\left\{12\sqrt{\frac{3}{(d+\tau)L}}+\sqrt{\frac{3\tilde{w}\alpha^2}{L}}\right\}\sqrt{\log(\frac{8N}{\epsilon})}$ and  $\lambda_{K}(\tilde{\Phi}_C^*(\alpha)Z^\top Z)>5 \alpha \varkappa/L+3\alpha\bar{c}/L$, then with probability at least $1-\epsilon$, there is a rotation matrix $\mathcal{O}$, so that 
\begin{equation}
    ||{U_{\Phi}-U_{\Phi}^*\mathcal{O}}||_F\leq \frac{16 K^{1/2}}{{\lambda_K(\tilde{\Phi}^*(\alpha, \tau))}}\frac{12\sqrt{3/(d+\tau)}+\sqrt{3{\tilde{w}}\alpha^2}}{\sqrt{L}}\sqrt{\log(\frac{8N}{\epsilon})}
\end{equation}
\end{proposition}

%-------------------------------------------------------------
To guarantee the convergence of eigenvectors, the smallest eigenvalue $\lambda_K$ to be sufficiently large. It is a common condition in matrix perturbation analysis; see \cite{davis1970, tong2023uniform}. 

Algorithm \ref{alg:sc} suggests that $k$-means clustering method based on $U_{\Phi}$ gives us the final labels. Let $C_{\Phi,i}, {C}_{\Phi,i}^*$ be the cluster centroid of the $i$th node generated using $k$-means clustering on $U_{\Phi}$ and $U_{\Phi}^*$ respectively. If rotation matrix $\mathcal{O}$ minimizes $\left\|{U_{\Phi}-U_{\Phi}^*\mathcal{O}}\right\|_{F}$, the set of mis-clustering nodes is: 
\[
\mathcal{M}=\{i: \mbox{ there exists $j \neq i$ such that }\|C_{\Phi,i} -{C}^*_{\Phi,i}\mathcal{O}\|_{2}>\|C_{\Phi,i}-{C}^*_{\Phi,j} \mathcal{O}\|_{2}\}.
\]
Under the definition of  mis-clustering nodes, Theorem \ref{thm:mis-SCL-N} bounds the mis-clustering rate $|\mathcal{M}|/N$.
\begin{theorem}
\label{thm:mis-SCL-N}
Let $\psi=\max_i (Z^\top Z)_{ii}$ denote the size of the largest block. Under the assumptions of the Proposition \ref{SCL-N-eigenvector}, with probability at least $1-\epsilon$, 
\begin{equation}
    \frac{|\mathcal{M}|}{N}\leq \frac{2048\psi K}{LN{\lambda_K^2(\tilde{\Phi}^*(\alpha, \tau))}}\biggl(12\sqrt{3/(d+\tau)}+\sqrt{3{\tilde{w}}\alpha^2}\biggr)^2\log(\frac{8N}{\epsilon})
\end{equation}
\end{theorem}

% {\color{blue} In the Theorem \ref{thm:mis-SCL-N}, it is in required of condition $(d+\tau)L>3\log(8N/\epsilon)$, which holds when $\sum_{l=1}^L\{p(l)+q(l)\}>\Theta \{\log (N)/N\}$ and results in $\sqrt{\frac{\log 8N/\epsilon}{L(d+\tau)}}=o(1)$, here we consider a node-contextualized multilayer stochastic block model with two multilayer-varying blocks, within block multilayer-varying probability $p(l)$, and between block multilayer-varying probability $q(l)$; it is also in required of condition $\tilde{w}>27N^2J^4\log(8N/\epsilon)/L$ holds when $R> \Theta(\log (N))$, which is a condition that requires the number of covariates to grow with the number of nodes. Here if we have each element of covariates matrix is bounded away from 0 uniformly, then $\alpha\leq \alpha_{\max}=\frac{\lambda_1(\tilde{\phi}^2_1)}{ \lambda_{R} (\tilde{H} )1_{R\leq K}+(\lambda_K( \tilde{H} )-\lambda_{K+1}(\tilde{H} ))1_{R> K}}=O(\frac{L}{NR})$ and $\tilde{w}=\Theta(\frac{N^2R}{L})$ by definition, we could infer $\sqrt{3\tilde{w}\alpha^2}=O(\sqrt{\frac{L}{R}})$. Thus we could get $\frac{1}{L}\left\{12\sqrt{3/(d+\tau)}+\sqrt{3\tilde{w}\alpha^2}\right\}^2\left\{\log(\frac{8N}{\epsilon})\right\}=o(1)$. Therefore, $\frac{|\mathcal{M}|}{N}=o(\frac{1}{\lambda_K(\tilde{\Phi}^*(\alpha, \tau))^2})$. 
% Above all, a larger rate of $L$, $d+\tau$ and $R$ would lower the bound of the mis-cluster rate. %It could also be seen that the rate would be lower as $L$ becomes larger. 

To better understand the complicated mis-clustering rate in Theorem \ref{thm:mis-SCL-N}, we consider a simplified model to illustrate the consistency of SCANC. 
Consider an MSBM-C with parameters $B([L]), M, J$ and $R > 1$. 
Suppose $B_{k_1k_2}(l)=p(l)$ and $B_{k_1k_2}(l)=q(l)$, $k_1 \neq k_2 \in [K]$. 
For the covariates, suppose $R$ is a multiple of $K$. 
The mean matrix follow that $M_{kr}=m_1$ when $r$ is a multiple of $k$ and $M_{kr}=m_2$ otherwise, $k\in [K], r\in [R]$. Hence, the mean covariate vector $M$ will have a repeating pattern. 
Suppose each community has the same number of nodes $N/K$. This is called a {\it simplified MSBM-C with parameters $p([L]), q([L]), m_1, m_2, J, R, K$.} 
The consistency of SCANC under the simplified MSBM-C model can be found in Corollary \ref{corollary}, where the detailed proof can be found in supplementary materials \cite{supp}.
\begin{corollary}
\label{corollary}
Consider a simplified MSBM-C model with parameters $p([L])$, $q([L])$, $ m_1$, $ m_2$, $ J$, $ R$, $ K$, where $K$ is fixed. 
Suppose for constants $a, b, c > 0$, the number of covariates $R=\Theta(L(\log N)^{a+1})$, the minimum expected degree $d+\tau=\Theta((\log N)^{b+1})$,  and the tuning parameter $\alpha=\Theta({(\log N)^{-(1+c)}/N})$, then the mis-clustering bound in Theorem \ref{thm:mis-SCL-N} becomes
\begin{equation}
 \frac{|\mathcal{M}|}{N}\leq\frac{C}{L}\frac{(\log N)^{-b}+(\log N)^{a-2c}+(\log N)^{a/2-c-b/2}}{\Theta(1)+(\log N)^{a-c}+(\log N)^{2a-2c}}
\end{equation}
where $C > 0$ is a constant. For $a<b$ the mis-clustering rate is minimized with $c=\frac{a+b}{2}$, which yields a rate of $\Theta({(\log N)^{-b}/L})$; for $a\geq b$ the mis-clustering rate is minimized with $c=0$, which yields a rate of $\Theta({(\log N)^{-a}/L})$. 
\end{corollary}

When the optimal tuning parameter $\alpha$ is selected in SCANC, the mis-clustering rate converges to 0 in the rate of $\Theta((\log N)^{\max\{a, b\}}/L)$. Here, $a$ gives the information level in covariates and $b$ gives the information level in network. Our method combines the information from covariates, network, and the number of layers $L$. 

\begin{remark}
Corollary \ref{corollary} also indicates that our data-driven selection criteria of the tuning parameter $\alpha$ is reasonable. When the networks contain more information, $\alpha$ should be selected to balance the network effects and covariate effects. When the covariates contain more information, then we should select a small $\alpha = O((N \log N)^{-1})$. When $\alpha$ is in the corresponding range, then SCANC yields the optimal rate. 
Consider the special case that $a = 0$ and $b = 0$. 
Our suggested region of $\alpha$ in Section \ref{subsec:alpha} follows that  $\alpha_{\max}=\alpha_{min}=\Theta((N\log N)^{-1})$. 
By Corollary \ref{corollary}, when $a = b$, the optimal choice should be  $\alpha = \Theta((N\log N)^{-1})$. This rate agrees with data-driven interval $[\alpha_{\min}, \alpha_{\max}]$. 
\end{remark}
%------------------------------------------------------------------------------------
\subsection{SCALC: Consistency, Tuning Parameters, and Comparison}\label{subsec:scalc}

In this section, we demonstrate the theoretical guarantee of SCALC. Since the proof is similar with that of SCANC, we skip the propositions and summarize the main results in Theorem \ref{thm:mis-SCL-N2}. We then discuss the upper bound under the simplified MSBM-C model in Corollary \ref{corollary2}. Finally, we compare the performance of SCANC and SCALC in Theorem \ref{thm:kth} to demonstrate the effects of network assortativity.  

To present the result, we need to define some terms. 
Let $d(l) = \min_i\E [D_{ii}(l)]$, $l \in [L]$. Hence, $d(l)$ is the minimum expected degree of layer $l$ and $d$ is the overall minimum. 
Define the community Laplacian matrix respective to SCALC: 
\begin{eqnarray*}
\tilde{\Psi}^*_C(\alpha) & = &\frac{1}{L}\sum_{l=1}^L(D_{C}^*(l)+\tau I)^{-1/2} B(l) Z^{\top}(D^*(l)+\tau I)^{-1}ZB(l)(D_{C}^*(l)+\tau I)^{-1/2}\\
\quad &&+\alpha MM^{\top}/L.
\end{eqnarray*}
\begin{theorem}
\label{thm:mis-SCL-N2}
Suppose for a constant $\epsilon > 0$, there is  $\frac{1}{L}\sum_{l=1}^L\frac{3\log(8N/\epsilon)}{d(l)+\tau}<1$, $\frac{3}{2}\frac{1}{L}\sum_{l=1}^L\frac{1}{d(l)/\tau+1}\geq 1$, $\tilde{w}>27N^2J^4\log(8N/\epsilon)/L$, %and $\lambda_{K}(\tilde{\Psi}^*_C(\alpha)Z^{\top} Z)>2 \alpha \varkappa/L$, 
$\lambda_K(\tilde{\Psi}^*(\alpha, \tau)) \geq \{(16+5\sqrt{6})\{\frac{3}{L}\sum_{l=1}^L\frac{1}{d(l)+\tau}\}^{1/2}+\sqrt{12\tilde{w}\alpha^2}\}\sqrt{\log(\frac{8N}{\epsilon})}/{\sqrt{L}}$, and $\lambda_{K}(\tilde{\Psi}^*_C(\alpha)Z^{\top} Z)>5 \alpha \varkappa/L+3\alpha\bar{c}/L$. Then with probability at least $1-\epsilon$, 
\begin{equation}
    \frac{|\mathcal{M}|}{N}\leq \frac{2048\psi K}{LN\lambda_K^2(\tilde{\Psi}^*(\alpha, \tau))}\biggl[(8+5\sqrt{6}/2)\{\frac{3}{L}\sum_{l=1}^L\frac{1}{d(l)+\tau}\}^{1/2}+\sqrt{3\tilde{w}\alpha^2}\biggr]^2 \log(\frac{8N}{\epsilon}).
\end{equation}
\end{theorem}

For SCALC, we have similar results on the tuning parameter $\alpha$ under the simplified MSBM-C model in Corollary \ref{corollary2}.  

\begin{corollary}
\label{corollary2}
Consider a simplified MSBM-C model with $p([L]), q([L]), m_1, m_2, J, R, K$, where $K$ is fixed and $R$ is a multiple of $K$. 
Suppose there are constants $a, b, c > 0$, and $R=\Theta\{L(\log N)^{a+1}\}$, $\frac{1}{L}\sum_{l=1}^L\frac{1}{d(l)+\tau}=\Theta\{(\log N)^{-b-1}\}$, and $\alpha=\Theta\{{(\log N)^{-1-c}}/N\}$, then the mis-clustering bound in Theorem \ref{thm:mis-SCL-N2} becomes
\begin{equation}
 \frac{|\mathcal{M}|}{N}\leq\frac{C}{L}\frac{(\log N)^{-b}+(\log N)^{a-2c}+(\log N)^{a/2-c-b/2}}{\Theta(1)+(\log N)^{a-c}+(\log N)^{2a-2c}}
\end{equation}
where $C > 0$ is a constant. For $a<b$ the mis-clustering rate is minimized with $c=\frac{a+b}{2}$, which yields a rate of $\Theta({(\log N)^{-b}}/L)$; for $a\geq b$ the mis-clustering rate is minimized with $c=0$, which yields a rate of $\Theta({(\log N)^{-a}/L})$. 
\end{corollary}

The assumptions and results of Theorems  \ref{thm:mis-SCL-N} and \ref{thm:mis-SCL-N2}  are very similar, since both are using Algorithm \ref{alg:sc} on the Laplacian matrix. 
One difference is that the term $\frac{1}{\min_{l \in [L]} d(l)+\tau}$ in Theorem \ref{thm:mis-SCL-N} is replaced by $\frac{1}{L} \sum_{l = 1}^L \frac{1}{d(l) + \tau}$ in Theorem \ref{thm:mis-SCL-N2}. Since $\frac{1}{\min_{l \in [L]} d(l)+\tau} \geq \frac{1}{L} \sum_{l = 1}^L \frac{1}{d(l) + \tau}$,  SCALC has a relaxed condition on the network at each layer. 
The other difference is the Laplacian matrices used, where Theorem \ref{thm:mis-SCL-N} employs $\tilde{\Phi}^*(\alpha, \tau)$ and $\tilde{\Phi}^*_C(\alpha)$, and Theorem \ref{thm:mis-SCL-N2} considers $\tilde{\Phi}^*(\alpha, \tau)$ and $\tilde{\Phi}^*_C(\alpha)$. 
The difference in their eigenvalues can be found in Theorem \ref{thm:kth} on the simplified MSBM-C model.

%it is easy to see that the mis-cluster rate of SCALC is larger than SCANC. However, SCALC is not influenced by the problem of assortative and disassortative networks, which could be seen from Theorem \ref{kth} as follows.
%Here following assumption of Corollary \ref{corollary}, we could get lower bound of $K$th eigenvalue of $\tilde{\mathcal{L}}^{*}(\alpha)$ in Theorem \ref{kth}. 

\begin{theorem}
\label{thm:kth}
Consider the simplified MSBM-C model with $p([L]), q([L]), m_1, m_2, J, R, K$. 
Further, suppose $p(l)+(K-1)q(l) = c_{p,q}$ is a fixed constant for all $l \in [L]$. 
The $K$-th eigenvalue of $\tilde{\Phi}^*(\alpha, \tau)$ and $\tilde{\Psi}^*(\alpha, \tau)$ follows:
\begin{eqnarray*}
 \lambda_K(\tilde{\Phi}^*(\alpha, \tau))
&\geq&\frac{\left(\frac{1}{L}\sum_{l=1}^L[p(l) - q(l)]\right)^2}{(c_{p,q}+K\tau/N)^2} +\frac{\alpha RN (m_1-m_2)^2}{LK^2},
\end{eqnarray*}
and
    \begin{eqnarray*}
 \lambda_K(\tilde{\Psi}^*(\alpha, \tau))
&\geq&\frac{\frac{1}{L}\sum_{l=1}^L[p(l)-q(l)]^2}{(c_{p,q}+K\tau/N)^2} +\frac{\alpha RN(m_1-m_2)^2}{LK^2}.
\end{eqnarray*}
\end{theorem}
The difference in the lower bound appears in the first term about networks. While the denominators are the same, the numerator in $\lambda_K(\tilde{\Phi}^*(\alpha, \tau))$ is $\left(\frac{1}{L}\sum_{l=1}^L[p(l) - q(l)]\right)^2$, squared of the average differences between the diagonals and off-diagonals; and the numerator in $\lambda_K(\tilde{\Psi}^*(\alpha, \tau))$ is $\frac{1}{L} \sum_{l=1}^L (p(l) - q(l))^2$, the average of squared differences. 
Therefore, SCANC will enjoy a larger $\lambda_K$ when $p(l) > q(l)$ (or $p(l) < q(l)$) holds for all $l \in [L]$ than SCALC. However, when there are both assortative networks and disassortative networks, then SCALC will enjoy a better performance by addressing the signal cancellation problem.

%------------------------------------------------------------------------------------
\section{Simulation}\label{sec:simulation}
We consider a multilayer network with covariates and perform two sets of numerical studies to evaluate the performance of our proposed approaches, focusing on their {consistency} as \( N \to \infty \) and {robustness} under mis-specification across layers. For each simulation setting, we examine two scenarios: 1) all layers exhibit assortative networks; and 2) both assortative and disassortative networks are present.

We consider a multilayer network with $N$ nodes, $L = 5$ layers, and $K = 3$ communities with equal community size. Let $z \in [K]^N$ denote the label vector, where $z_i$ gives the label of node $i$. 
For the $l$-th layer, the network $W_{ij}(l) \sim Bernoulli(B_{z_i, z_j}(l))$ independently. 
The matrix $B \in \reals^{K \times K}$ is defined as:  
\begin{equation}\label{e1}
B_{k_1, k_2}^{(1)}(l) = \left\{\begin{array}{ll}
0.005 l + 0.015, & k_1 = k_2,\\
0.015, & k_1 \neq k_2,\\
\end{array}
\right. \quad 
B_{k_1, k_2}^{(2)}(l) = \left\{\begin{array}{ll}
0.045, &  k_1 = k_2,\\
0.01 l, &  k_1 \neq k_2.\\
\end{array}
\right. 
\end{equation}
Here, $B^{(1)}([L])$ presents all-assortative multilayer networks, and $B^{(2)}([L])$ indicates a scenario that the network is assortative on Layers 1, 2, 3, 4, but disassortative on Layer 5. 
To facilitate methods comparison, the covariates are generated differently for each layer. Let $Y_i(l) \sim N((-0.4+0.3 l) e_{z_i}, I_3)$ be the covariate vector on $l$-th layer for node $i$, where $e_{z_i} \in \reals^3$ has 1 on $z_i$-th entry and 0 on other entries. 
A larger $l$ indicates a stronger signal in the covariates. 

Two sets of approaches are compared: 1) methods only based on the network $W([L])$, including the bias-corrected spectral clustering approach (BC-Adj) in \cite{lei2022biasadjusted} and the multiple adjacency spectral embedding approach (MASE) in \cite{arroyo2021inference}; 2) methods based on both the networks $W([L])$ and covariates $Y$, including the tensor approach on covariate-assisted multilayer stochastic blockmodel (CAMSBM) in \cite{xu2022covariate}, the dynamic covariate-assisted spectral clustering approach (CASC) in \cite{guo2022time}, the least square method (LSE) in \cite{lei2020}, and the spectral clustering on mean adjacency matrix approach (MeanAdj) in \cite{han2015}. 
Here, CASC considers layer-variant covariates and CAMSBM uses the covariate matrix $Y = [Y(1), Y(2), \cdots, Y(L)]$, where $Y(l)$ is the covariate matrix consisting of $Y_i(l)$ as rows. 
We want to remind the readers that LSE and MeanAdj are actually methods on only networks. Following the suggestion in their papers, we build $L$ additional layers based on the similarity of covariate matrix in each layer, and then apply their methods.

For any estimated label vector $\hat{z}$, the mis-clustering rate is defined as 
\begin{equation}\label{eqn:err}
    Err(\hat{z}) = \min_{\pi: [K] \to [K]} \frac{1}{N} \sum_{i=1}^N 1_{z_i \neq \pi(\hat{z}_{i})}. 
\end{equation}
It is used to evaluate the performance of methods.

\begin{figure}[!htbp]
 \includegraphics[width = 1\textwidth]{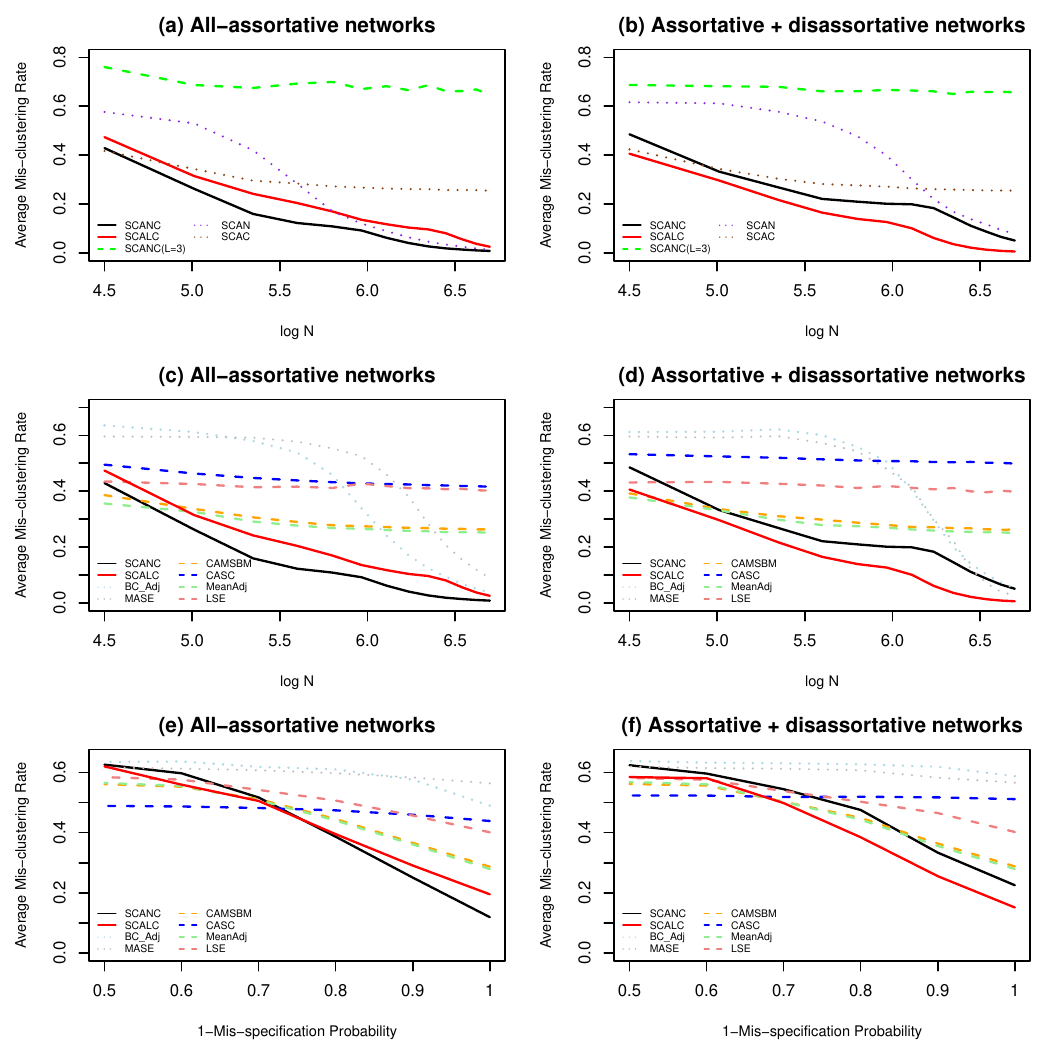}
  \caption{Mis-clustering rate of our approach and other approaches. Top row: Experiment Ia with spectral clustering methods. Second row: Experiment Ib. Third row: Experiment II.}
   \label{exp1}
\end{figure}

{\bf Experiment 1a}. We examine the consistency of spectral clustering approaches based on networks and covariates. Let $N \in \{90, 150, 210,\cdots, 810\}$. 
For Algorithm \ref{alg:sc}, we take the input matrix $A$ as aggregated networks and covariates (SCANC), aggregated Laplacians and covariates (SCALC), aggregated networks (SCAN), covariates (SCAC), and networks and covariates on a single layer $l = 3$. 
The average mis-clustering rates over 100 repetitions are summarized in Figure \ref{exp1}(a) and (b). 

The mis-clustering rates all decrease to 0 as $N$ increases, demonstrating the consistency of all methods.
SCANC and SCALC outperform all other methods, which proves the effects of aggregation on covariates, networks, and layers. Further, SCANC works better in the all-assortative case and SCALC works better in the assortative-and-disassortative case, as we pointed out.

{\bf Experiment 1b}. On the same setting that $N \in \{90, 150, 210,\cdots, 810\}$, we compare our approach with other multilayer network community detection approaches. The average mis-clustering rates over 100 repetitions are summarized in Figure \ref{exp1}(c) and (d). 
Our newly proposed SCANC and SCALC methods outperform all other approaches. The network-based methods, MASE and BC-Adj highly depend on $N$. Compared to them, methods based on both network and covariates are less sensitive, but highly rely on the covariates. Our method combines the signal from both parts.

{\bf Experiment 2}. We explore the robustness when nodes have layer-varying labels. Let $N = 300$. Let $z$ be the overall community label vector and $z(l)$ be the label vector of the $l$-th layer. 
For any $i \in [N]$ and $l \in [L]$, $z_i(l) = z_i$ with probability $1 - q$ and $z_i(l) \in [K]/\{z_i\}$ with probability $q$. We call $q$ as the mis-specification rate. Based on $z(l)$, we generate the networks $W([L])$ and covariates $Y([L])$ the same way as Experiment 1a. Our goal is to recover the overall label vector $z$.

The average mis-clustering rate over 100 repetitions for all methods are summarized in Fig. \ref{exp1} (e) and (f). All methods fail when $q$ approaches 0.5. As the mis-specification rate decreases, methods gradually improve. Among them, our newly proposed SCANC and SCALC methods outperform all others, which proves its robustness in the presence of mis-specification.

\section{Primary School Contact Data Analysis}\label{sec:data}

The high-resolution contact network dataset was collected over two days in October 2009 at a primary school in Lyon, France, as detailed by \cite{stehle2011high}. 
Students wear devices to record their real-time interactions in each 20 seconds. 
For simplification, we divide one school day into 17 time intervals according to the activities, each lasting about 25 minutes. 
For one time interval, we build a network where each node represents a student and each edge represents non-zero interactions between them during this time interval. 
Therefore, we construct a multilayer network with $L = 34$ on two days for each grade. 

The covariates are constructed based on the gender-specific interactions during each time interval. 
Let $Y = [Y(1), Y(2), \cdots, Y(34)]$. Here, $Y(l) \in \reals^{N \times 14}$ is the covariates matrix based on interactions in $l$-th time interval. 
In detail, for student $i$, $Y_i(l)$ records the proportion of interactions with a given gender that span exactly $m$ consecutive $20$-second intervals among all his/her interactions, where $m = 1, 2, 3, 5, 6$, or $\geq 7$. In total, there are 476 covariates across all layers.

We first interpret our community detection results using Grade 5 students, and then compare the clustering accuracy with other methods across all grades. 
Among the $45$ students in Grade 5, 21 belong to class 5A and the other 24 are in class 5B. The interactions highly rely on the class division and partially rely on genders. 

The result of our newly proposed SCANC algorithm when $K = 4$ is presented in Table \ref{scancgrade}. We choose SCANC instead of SCALC, since there should always be more interactions between students in the same community, i.e. the same network assortativity among layers. Due to the class activities, the two classes can be almost clearly divided by clusters 1/2 and 3/4, with only one error. 
Further, clusters 3 and 4 perfectly cluster the students in 5B by their gender, while clusters 1 and 2 have very few errors. A close investigation of the dataset in supplementary materials suggests that: 1) the class-based division is primarily shaped by the network; and 2) the covariates contain gender information. Our SCANC method leverages the information from both sides to form a clear division. 
\begin{table}[htbp] 
		\caption{Estimated  communities of SCANC on Grade 5 students} \label{scancgrade}
        \centering
		\begin{tabular}{@{}l|llll|l@{}}
\toprule
    Community &   5A female & 5A male & 5B female & 5B male  & Total\\
\midrule
     1&7&4&0&0&11\\
     2&3&6&0&0&9\\
     3&0& 0&11&0&11\\
     4&0&1&0&13&14\\
     \hline
     Total & 10 & 11 & 11 & 13 & 45\\
\botrule
		\end{tabular}
\end{table}

We then compare our newly proposed approaches with other methods on all 5 datasets. 
For each method, we evaluate the community detection performance by the Normalized Mutual Information (NMI) between the estimated labels and the labels naturally given by classes and genders. A higher NMI indicates a more consistent clustering. 
Since there are too few nodes with unknown gender, we merge them with the other groups and let $K = 4$. 
For each estimated label vector, we report the NMI among all possible mergings to evaluate this estimate. 

The results are summarized in Table \ref{realdataresult}, where SCANC achieves the highest NMI in 4 out of 5 datasets. For Grade 4, the interactions between male and female groups in Class 4B are as frequent as those within groups; see Supplementary Materials. Hence, we further consider $K = 3$ for Grade 4, and let the true labels be 4A male, 4A female, and 4B. In this case, our method outperforms all other methods again.

\begin{table}[htbp] %[htbp] % [H]
		\caption{$N$, number of nodes; $K$, number of community. For each method, the NMI value is recorded.} \label{realdataresult}
		\begin{tabular}{@{}l|ll|ll|llll|ll@{}}
    \toprule
  Data &  $N$&$K$&SCANC&SCALC&CASC&CAMSBM&LSE&MeanAdj&BC-Adj&MASE\\
        \midrule
     Grade 1&46&4& {\bf.619}&.535&.544&.596&.529&.538&.515&.532\\
     Grade 2&49&4&{\bf.652}&.619&.621&.649&.640&.560&.594&.582\\
     Grade 3&45&4&{\bf.619} &.405 &.447&.506 &.469&.530 &.517 &.463  \\
    Grade 4&42&4&.552 &.491 &.507 &.614 &{\bf .650} &.585 &.546 &.593\\
     Grade 4&42&3\footnotemark[1]&{\bf.711} &.671 &.669 &.473 &{\bf.711} &.675 &.671 &{\bf.711}\\
     Grade 5&45&4&{\bf.736} &.455 &.683 &.675 &.669 &.512 &.417 &.515\\
			\botrule
		\end{tabular}
		\footnotetext[1]{Another division of Grade 4: Class 4A male, Class 4A female, and Class 4B.}
\end{table}

%----------------------------------------------------------------
%----------------------------------------------------------------
%----------------------------------------------------------------

 \section{Discussion}\label{sec:disc}
In this paper, we proposed two spectral-based methods, Spectral Clustering on Aggregated Networks with Covariates (SCANC) and Spectral Clustering on Aggregated Laplacian with Covariates (SCALC), to address the problem of community detection in multilayer networks with covariates. Our theoretical analysis under the Multilayer Stochastic Blockmodel with Covariates (MSBM-C) demonstrated the consistency of both methods, with error bounds derived based on the combined signal-to-noise ratio of the networks and covariates across layers. The simulation studies and real-data application further validated the effectiveness of our approaches.

Our work demonstrate three key findings. 
First, our results confirm that aggregating network layers generally improves clustering performance by increasing the signal-to-noise ratio. SCANC is particularly effective when all layers exhibit consistent assortativity, while SCALC is advantageous in scenarios where both assortative and disassortative layers are present, as it mitigates signal cancellation issues. 
Second, compared to methods that rely solely on the network structure, our approaches leverage nodal covariates, which provide additional information for community detection. Both theoretical and numerical results indicate that leveraging nodal covariates significantly enhance clustering accuracy.
Finally, tuning parameter selection is crucial but robust in our algorithm.
The weight parameter $\alpha$, which balances the influence of network and covariates, is shown to play a pivotal role in clustering performance. Our data-driven selection approach ensures that $\alpha$ remains within an optimal range, leading to stable and reliable community detection results.

There are many open questions to explore in this area.
Our approaches rely on linear aggregation of adjacency matrices or Laplacians. Exploring alternative nonlinear or adaptive aggregation strategies, potentially leveraging deep learning techniques, could further enhance performance.
Real-world networks often contain missing or corrupted covariate information. Investigating robustness under partial covariate observation or developing imputation strategies would be valuable extensions of this work.

% {\color{blue} 
% In this paper, we propose two aggregated spectral methods for community detection in multilayer networks with covariates: Spectral Clustering on Aggregated Networks and Covariates (SCANC) and Spectral Clustering on Aggregated Network Laplacian and Covariates (SCALC). We derive asymptotic results for the mis-clustering rate, analyze tuning parameters, and compare the performance of SCANC and SCALC, supported by simulations.

% SCANC is most effective when the networks in the multilayer model are consistent and assortative. However, when both assortative and disassortative networks are present, SCALC outperforms SCANC.}

%\bibliographystyle{Chicago}

\section*{Supplementary material}
\label{SM}

The Supplementary material  includes additional results on the primary school contact network dataset, theoretical proofs of theorems and additional theoretical results.

\section*{Acknowledgements}
This research was supported by Singapore Ministry of Education Academic Research Fund Tier 1 A-00004813-00-00 and the National University of Singapore Scholarship.

\bibliography{sample}


\begin{thebibliography}{40}
\ifx \bisbn   \undefined \def \bisbn  #1{ISBN #1}\fi
\ifx \binits  \undefined \def \binits#1{#1}\fi
\ifx \bauthor  \undefined \def \bauthor#1{#1}\fi
\ifx \batitle  \undefined \def \batitle#1{#1}\fi
\ifx \bjtitle  \undefined \def \bjtitle#1{#1}\fi
\ifx \bvolume  \undefined \def \bvolume#1{\textbf{#1}}\fi
\ifx \byear  \undefined \def \byear#1{#1}\fi
\ifx \bissue  \undefined \def \bissue#1{#1}\fi
\ifx \bfpage  \undefined \def \bfpage#1{#1}\fi
\ifx \blpage  \undefined \def \blpage #1{#1}\fi
\ifx \burl  \undefined \def \burl#1{\textsf{#1}}\fi
\ifx \doiurl  \undefined \def \doiurl#1{\url{https://doi.org/#1}}\fi
\ifx \betal  \undefined \def \betal{\textit{et al.}}\fi
\ifx \binstitute  \undefined \def \binstitute#1{#1}\fi
\ifx \binstitutionaled  \undefined \def \binstitutionaled#1{#1}\fi
\ifx \bctitle  \undefined \def \bctitle#1{#1}\fi
\ifx \beditor  \undefined \def \beditor#1{#1}\fi
\ifx \bpublisher  \undefined \def \bpublisher#1{#1}\fi
\ifx \bbtitle  \undefined \def \bbtitle#1{#1}\fi
\ifx \bedition  \undefined \def \bedition#1{#1}\fi
\ifx \bseriesno  \undefined \def \bseriesno#1{#1}\fi
\ifx \blocation  \undefined \def \blocation#1{#1}\fi
\ifx \bsertitle  \undefined \def \bsertitle#1{#1}\fi
\ifx \bsnm \undefined \def \bsnm#1{#1}\fi
\ifx \bsuffix \undefined \def \bsuffix#1{#1}\fi
\ifx \bparticle \undefined \def \bparticle#1{#1}\fi
\ifx \barticle \undefined \def \barticle#1{#1}\fi
\bibcommenthead
\ifx \bconfdate \undefined \def \bconfdate #1{#1}\fi
\ifx \botherref \undefined \def \botherref #1{#1}\fi
\ifx \url \undefined \def \url#1{\textsf{#1}}\fi
\ifx \bchapter \undefined \def \bchapter#1{#1}\fi
\ifx \bbook \undefined \def \bbook#1{#1}\fi
\ifx \bcomment \undefined \def \bcomment#1{#1}\fi
\ifx \oauthor \undefined \def \oauthor#1{#1}\fi
\ifx \citeauthoryear \undefined \def \citeauthoryear#1{#1}\fi
\ifx \endbibitem  \undefined \def \endbibitem {}\fi
\ifx \bconflocation  \undefined \def \bconflocation#1{#1}\fi
\ifx \arxivurl  \undefined \def \arxivurl#1{\textsf{#1}}\fi
\csname PreBibitemsHook\endcsname

%%% 1
\bibitem[\protect\citeauthoryear{Chen and Yuan}{2006}]{chen2006detecting}
\begin{barticle}
\bauthor{\bsnm{Chen}, \binits{J.}},
\bauthor{\bsnm{Yuan}, \binits{B.}}:
\batitle{Detecting functional modules in the yeast protein--protein interaction
  network}.
\bjtitle{Bioinformatics}
\bvolume{22}(\bissue{18}),
\bfpage{2283}--\blpage{2290}
(\byear{2006})
\end{barticle}
\endbibitem

%%% 2
\bibitem[\protect\citeauthoryear{Sporns and Betzel}{2016}]{sporns2016modular}
\begin{barticle}
\bauthor{\bsnm{Sporns}, \binits{O.}},
\bauthor{\bsnm{Betzel}, \binits{R.F.}}:
\batitle{Modular brain networks}.
\bjtitle{Annual Review of Psychology}
\bvolume{67},
\bfpage{613}
(\byear{2016})
\end{barticle}
\endbibitem

%%% 3
\bibitem[\protect\citeauthoryear{Ying et~al.}{2018}]{ying2018graph}
\begin{botherref}
\oauthor{\bsnm{Ying}, \binits{R.}},
\oauthor{\bsnm{He}, \binits{R.}},
\oauthor{\bsnm{Chen}, \binits{K.}},
\oauthor{\bsnm{Eksombatchai}, \binits{P.}},
\oauthor{\bsnm{Hamilton}, \binits{W.L.}},
\oauthor{\bsnm{Leskovec}, \binits{J.}}:
Graph convolutional neural networks for web-scale recommender systems.
Proceedings of the 24th ACM SIGKDD International Conference on Knowledge
  Discovery and Data Mining,
974--983
(2018)
\end{botherref}
\endbibitem

%%% 4
\bibitem[\protect\citeauthoryear{Leskovec and
  Mcauley}{2012}]{leskovec2012learning}
\begin{botherref}
\oauthor{\bsnm{Leskovec}, \binits{J.}},
\oauthor{\bsnm{Mcauley}, \binits{J.}}:
Learning to discover social circles in ego networks.
Advances in Neural Information Processing Systems
\textbf{25}
(2012)
\end{botherref}
\endbibitem

%%% 5
\bibitem[\protect\citeauthoryear{De~Bacco et~al.}{2017}]{de2017community}
\begin{barticle}
\bauthor{\bsnm{De~Bacco}, \binits{C.}},
\bauthor{\bsnm{Power}, \binits{E.A.}},
\bauthor{\bsnm{Larremore}, \binits{D.B.}},
\bauthor{\bsnm{Moore}, \binits{C.}}:
\batitle{Community detection, link prediction, and layer interdependence in
  multilayer networks}.
\bjtitle{Physical Review E}
\bvolume{95}(\bissue{4}),
\bfpage{042317}
(\byear{2017})
\end{barticle}
\endbibitem

%%% 6
\bibitem[\protect\citeauthoryear{Jing et~al.}{2021}]{jing2021community}
\begin{barticle}
\bauthor{\bsnm{Jing}, \binits{B.-Y.}},
\bauthor{\bsnm{Li}, \binits{T.}},
\bauthor{\bsnm{Lyu}, \binits{Z.}},
\bauthor{\bsnm{Xia}, \binits{D.}}:
\batitle{Community detection on mixture multilayer networks via regularized
  tensor decomposition}.
\bjtitle{The Annals of Statistics}
\bvolume{49}(\bissue{6}),
\bfpage{3181}--\blpage{3205}
(\byear{2021})
\end{barticle}
\endbibitem

%%% 7
\bibitem[\protect\citeauthoryear{Hu et~al.}{2022}]{hu2022graph}
\begin{barticle}
\bauthor{\bsnm{Hu}, \binits{Y.}},
\bauthor{\bsnm{Wang}, \binits{W.}},
\bauthor{\bsnm{Yu}, \binits{Y.}}:
\batitle{Graph matching beyond perfectly-overlapping erd{\H{o}}s--r{\'e}nyi
  random graphs}.
\bjtitle{Statistics and Computing}
\bvolume{32}(\bissue{1}),
\bfpage{19}
(\byear{2022})
\end{barticle}
\endbibitem

%%% 8
\bibitem[\protect\citeauthoryear{Ma and Nandy}{2023}]{ma2023community}
\begin{barticle}
\bauthor{\bsnm{Ma}, \binits{Z.}},
\bauthor{\bsnm{Nandy}, \binits{S.}}:
\batitle{Community detection with contextual multilayer networks}.
\bjtitle{IEEE Transactions on Information Theory}
\bvolume{69}(\bissue{5}),
\bfpage{3203}--\blpage{3239}
(\byear{2023})
\end{barticle}
\endbibitem

%%% 9
\bibitem[\protect\citeauthoryear{Binkiewicz
  et~al.}{2017}]{binkiewicz2017covariate}
\begin{barticle}
\bauthor{\bsnm{Binkiewicz}, \binits{N.}},
\bauthor{\bsnm{Vogelstein}, \binits{J.T.}},
\bauthor{\bsnm{Rohe}, \binits{K.}}:
\batitle{Covariate-assisted spectral clustering}.
\bjtitle{Biometrika}
\bvolume{104}(\bissue{2}),
\bfpage{361}--\blpage{377}
(\byear{2017})
\end{barticle}
\endbibitem

%%% 10
\bibitem[\protect\citeauthoryear{Yan and Sarkar}{2021}]{yan2021covariate}
\begin{barticle}
\bauthor{\bsnm{Yan}, \binits{B.}},
\bauthor{\bsnm{Sarkar}, \binits{P.}}:
\batitle{Covariate regularized community detection in sparse graphs}.
\bjtitle{Journal of the American Statistical Association}
\bvolume{116}(\bissue{534}),
\bfpage{734}--\blpage{745}
(\byear{2021})
\end{barticle}
\endbibitem

%%% 11
\bibitem[\protect\citeauthoryear{Zhang et~al.}{2016}]{zhang2015community}
\begin{barticle}
\bauthor{\bsnm{Zhang}, \binits{Y.}},
\bauthor{\bsnm{Levina}, \binits{E.}},
\bauthor{\bsnm{Zhu}, \binits{J.}}:
\batitle{Community detection in networks with node features}.
\bjtitle{Electronic Journal of Statistics}
\bvolume{10}(\bissue{2}),
\bfpage{3153}
(\byear{2016})
\end{barticle}
\endbibitem

%%% 12
\bibitem[\protect\citeauthoryear{Hu and Wang}{2024}]{hu2024network}
\begin{barticle}
\bauthor{\bsnm{Hu}, \binits{Y.}},
\bauthor{\bsnm{Wang}, \binits{W.}}:
\batitle{Network-adjusted covariates for community detection}.
\bjtitle{Biometrika}
\bvolume{111}(\bissue{4}),
\bfpage{1221}--\blpage{1240}
(\byear{2024})
\end{barticle}
\endbibitem

%%% 13
\bibitem[\protect\citeauthoryear{Nowicki and Snijders}{2001}]{nowicki2001}
\begin{barticle}
\bauthor{\bsnm{Nowicki}, \binits{K.}},
\bauthor{\bsnm{Snijders}, \binits{T.A.B.}}:
\batitle{Estimation and prediction for stochastic blockstructures}.
\bjtitle{Journal of the American Statistical Association}
\bvolume{96}(\bissue{455}),
\bfpage{1077}--\blpage{1087}
(\byear{2001})
\end{barticle}
\endbibitem

%%% 14
\bibitem[\protect\citeauthoryear{Airoldi et~al.}{2008}]{airoldi2008}
\begin{barticle}
\bauthor{\bsnm{Airoldi}, \binits{E.M.}},
\bauthor{\bsnm{Blei}, \binits{D.M.}},
\bauthor{\bsnm{Fienberg}, \binits{S.E.}},
\bauthor{\bsnm{Xing}, \binits{E.P.}}:
\batitle{Mixed membership stochastic blockmodels}.
\bjtitle{Journal of Machine Learning Research}
\bvolume{9},
\bfpage{1981}--\blpage{2014}
(\byear{2008})
\end{barticle}
\endbibitem

%%% 15
\bibitem[\protect\citeauthoryear{Handcock et~al.}{2007}]{handcock2007}
\begin{barticle}
\bauthor{\bsnm{Handcock}, \binits{M.S.}},
\bauthor{\bsnm{Raftery}, \binits{A.E.}},
\bauthor{\bsnm{Tantrum}, \binits{J.M.}}:
\batitle{Model-based clustering for social networks}.
\bjtitle{Journal of the Royal Statistical Society: Series A (Statistics in
  Society)}
\bvolume{170}(\bissue{2}),
\bfpage{301}--\blpage{354}
(\byear{2007})
\end{barticle}
\endbibitem

%%% 16
\bibitem[\protect\citeauthoryear{Amini et~al.}{2013}]{amini2013}
\begin{barticle}
\bauthor{\bsnm{Amini}, \binits{A.A.}},
\bauthor{\bsnm{Chen}, \binits{A.}},
\bauthor{\bsnm{Bickel}, \binits{P.J.}},
\bauthor{\bsnm{Levina}, \binits{E.}}:
\batitle{Pseudo-likelihood methods for community detection in large sparse
  networks}.
\bjtitle{The Annals of Statistics}
\bvolume{41}(\bissue{4}),
\bfpage{2097}--\blpage{2122}
(\byear{2013})
\end{barticle}
\endbibitem

%%% 17
\bibitem[\protect\citeauthoryear{Chaudhuri et~al.}{2012}]{2012spectral}
\begin{barticle}
\bauthor{\bsnm{Chaudhuri}, \binits{K.}},
\bauthor{\bsnm{Chung}, \binits{F.}},
\bauthor{\bsnm{Tsiatas}, \binits{A.}}:
\batitle{Spectral clustering of graphs with general degrees in the extended
  planted partition model}.
\bjtitle{Journal of Machine Learning Research}
\bvolume{23},
\bfpage{1}--\blpage{23}
(\byear{2012})
\end{barticle}
\endbibitem

%%% 18
\bibitem[\protect\citeauthoryear{Qin and Rohe}{2013}]{qin2013}
\begin{botherref}
\oauthor{\bsnm{Qin}, \binits{T.}},
\oauthor{\bsnm{Rohe}, \binits{K.}}:
Regularized spectral clustering under the degree-corrected stochastic
  blockmodel.
Advances in Neural Information Processing Systems
\textbf{26}
(2013)
\end{botherref}
\endbibitem

%%% 19
\bibitem[\protect\citeauthoryear{Joseph and Yu}{2016}]{joseph2016impact}
\begin{botherref}
\oauthor{\bsnm{Joseph}, \binits{A.}},
\oauthor{\bsnm{Yu}, \binits{B.}}:
Impact of regularization on spectral clustering.
The Annals of Statistics,
1765--1791
(2016)
\end{botherref}
\endbibitem

%%% 20
\bibitem[\protect\citeauthoryear{Jin}{2015}]{jin2015}
\begin{barticle}
\bauthor{\bsnm{Jin}, \binits{J.}}:
\batitle{Fast community detection by {SCORE}}.
\bjtitle{The Annals of Statistics}
\bvolume{43}(\bissue{1}),
\bfpage{57}--\blpage{89}
(\byear{2015})
\end{barticle}
\endbibitem

%%% 21
\bibitem[\protect\citeauthoryear{Han et~al.}{2015}]{han2015}
\begin{bchapter}
\bauthor{\bsnm{Han}, \binits{Q.}},
\bauthor{\bsnm{Xu}, \binits{K.}},
\bauthor{\bsnm{Airoldi}, \binits{E.}}:
\bctitle{Consistent estimation of dynamic and multi-layer block models}.
In: \bbtitle{International Conference on Machine Learning},
pp. \bfpage{1511}--\blpage{1520}
(\byear{2015}).
\bcomment{PMLR}
\end{bchapter}
\endbibitem

%%% 22
\bibitem[\protect\citeauthoryear{Paul and Chen}{2020}]{paul2020spectral}
\begin{barticle}
\bauthor{\bsnm{Paul}, \binits{S.}},
\bauthor{\bsnm{Chen}, \binits{Y.}}:
\batitle{Spectral and matrix factorization methods for consistent community
  detection in multi-layer networks}.
\bjtitle{The Annals of Statistics}
\bvolume{48}(\bissue{1}),
\bfpage{230}--\blpage{250}
(\byear{2020})
\end{barticle}
\endbibitem

%%% 23
\bibitem[\protect\citeauthoryear{Bhattacharyya and
  Chatterjee}{2017}]{bhattacharyya2018spectral}
\begin{barticle}
\bauthor{\bsnm{Bhattacharyya}, \binits{S.}},
\bauthor{\bsnm{Chatterjee}, \binits{S.}}:
\batitle{Spectral clustering for multiple sparse networks: I}.
\bjtitle{Biometrika}
\bvolume{103}(\bissue{1}),
\bfpage{1}--\blpage{28}
(\byear{2017})
\end{barticle}
\endbibitem

%%% 24
\bibitem[\protect\citeauthoryear{Lei and Lin}{2023}]{lei2022biasadjusted}
\begin{barticle}
\bauthor{\bsnm{Lei}, \binits{J.}},
\bauthor{\bsnm{Lin}, \binits{K.Z.}}:
\batitle{Bias-adjusted spectral clustering in multi-layer stochastic block
  models}.
\bjtitle{Journal of the American Statistical Association}
\bvolume{118}(\bissue{544}),
\bfpage{2433}--\blpage{2445}
(\byear{2023})
\end{barticle}
\endbibitem

%%% 25
\bibitem[\protect\citeauthoryear{MacDonald et~al.}{2022}]{macdonald2022latent}
\begin{barticle}
\bauthor{\bsnm{MacDonald}, \binits{P.W.}},
\bauthor{\bsnm{Levina}, \binits{E.}},
\bauthor{\bsnm{Zhu}, \binits{J.}}:
\batitle{Latent space models for multiplex networks with shared structure}.
\bjtitle{Biometrika}
\bvolume{109}(\bissue{3}),
\bfpage{683}--\blpage{706}
(\byear{2022})
\end{barticle}
\endbibitem

%%% 26
\bibitem[\protect\citeauthoryear{Lyu et~al.}{2023}]{lyu2022}
\begin{barticle}
\bauthor{\bsnm{Lyu}, \binits{Z.}},
\bauthor{\bsnm{Xia}, \binits{D.}},
\bauthor{\bsnm{Zhang}, \binits{Y.}}:
\batitle{Latent space model for higher-order networks and generalized tensor
  decomposition}.
\bjtitle{Journal of Computational and Graphical Statistics}
\bvolume{32}(\bissue{4}),
\bfpage{1320}--\blpage{1336}
(\byear{2023})
\end{barticle}
\endbibitem

%%% 27
\bibitem[\protect\citeauthoryear{Contisciani et~al.}{2020}]{contisciani2020}
\begin{barticle}
\bauthor{\bsnm{Contisciani}, \binits{M.}},
\bauthor{\bsnm{Power}, \binits{E.A.}},
\bauthor{\bsnm{De~Bacco}, \binits{C.}}:
\batitle{Community detection with node attributes in multilayer networks}.
\bjtitle{Scientific Reports}
\bvolume{10}(\bissue{1}),
\bfpage{1}--\blpage{16}
(\byear{2020})
\end{barticle}
\endbibitem

%%% 28
\bibitem[\protect\citeauthoryear{Xu et~al.}{2022}]{xu2022covariate}
\begin{botherref}
\oauthor{\bsnm{Xu}, \binits{S.}},
\oauthor{\bsnm{Zhen}, \binits{Y.}},
\oauthor{\bsnm{Wang}, \binits{J.}}:
Covariate-assisted community detection in multi-layer networks.
Journal of Business and Economic Statistics,
1--12
(2022)
\end{botherref}
\endbibitem

%%% 29
\bibitem[\protect\citeauthoryear{Guo et~al.}{2024}]{guo2022time}
\begin{barticle}
\bauthor{\bsnm{Guo}, \binits{L.}},
\bauthor{\bsnm{H{\"a}rdle}, \binits{W.K.}},
\bauthor{\bsnm{Tao}, \binits{Y.}}:
\batitle{A time-varying network for cryptocurrencies}.
\bjtitle{Journal of Business and Economic Statistics}
\bvolume{42}(\bissue{2}),
\bfpage{437}--\blpage{456}
(\byear{2024})
\end{barticle}
\endbibitem

%%% 30
\bibitem[\protect\citeauthoryear{Lee et~al.}{2010}]{Lee}
\begin{barticle}
\bauthor{\bsnm{Lee}, \binits{A.B.}},
\bauthor{\bsnm{Luca}, \binits{D.}},
\bauthor{\bsnm{Roeder}, \binits{K.}}:
\batitle{A spectral graph approach to discovering genetic ancestry}.
\bjtitle{The Annals of Applied Statistics}
\bvolume{4}(\bissue{1}),
\bfpage{179}--\blpage{202}
(\byear{2010})
\end{barticle}
\endbibitem

%%% 31
\bibitem[\protect\citeauthoryear{Jin et~al.}{2017}]{jin2017phase}
\begin{barticle}
\bauthor{\bsnm{Jin}, \binits{J.}},
\bauthor{\bsnm{Ke}, \binits{Z.T.}},
\bauthor{\bsnm{Wang}, \binits{W.}}:
\batitle{Phase transitions for high dimensional clustering and related
  problems}.
\bjtitle{The Annals of Statistics}
\bvolume{45}(\bissue{5}),
\bfpage{2151}--\blpage{2189}
(\byear{2017})
\end{barticle}
\endbibitem

%%% 32
\bibitem[\protect\citeauthoryear{Zhao et~al.}{2025}]{supp}
\begin{botherref}
\oauthor{\bsnm{Zhao}, \binits{D.}},
\oauthor{\bsnm{Wang}, \binits{W.}},
\oauthor{\bsnm{Li}, \binits{J.}}:
Supplementary materials for ``spectral clustering on multilayer networks with
  covariates".
Manuscript
(2025)
\end{botherref}
\endbibitem

%%% 33
\bibitem[\protect\citeauthoryear{Holland et~al.}{1983}]{holland1983}
\begin{barticle}
\bauthor{\bsnm{Holland}, \binits{P.W.}},
\bauthor{\bsnm{Laskey}, \binits{K.B.}},
\bauthor{\bsnm{Leinhardt}, \binits{S.}}:
\batitle{Stochastic blockmodels: First steps}.
\bjtitle{Social Networks}
\bvolume{5}(\bissue{2}),
\bfpage{109}--\blpage{137}
(\byear{1983})
\end{barticle}
\endbibitem

%%% 34
\bibitem[\protect\citeauthoryear{Lei and Rinaldo}{2015}]{Lei2015}
\begin{barticle}
\bauthor{\bsnm{Lei}, \binits{J.}},
\bauthor{\bsnm{Rinaldo}, \binits{A.}}:
\batitle{Consistency of spectral clustering in stochastic block models}.
\bjtitle{The Annals of Statistics}
\bvolume{43}(\bissue{1}),
\bfpage{215}--\blpage{237}
(\byear{2015})
\end{barticle}
\endbibitem

%%% 35
\bibitem[\protect\citeauthoryear{Chen et~al.}{2022}]{chen2022global}
\begin{barticle}
\bauthor{\bsnm{Chen}, \binits{S.}},
\bauthor{\bsnm{Liu}, \binits{S.}},
\bauthor{\bsnm{Ma}, \binits{Z.}}:
\batitle{Global and individualized community detection in inhomogeneous
  multilayer networks}.
\bjtitle{The Annals of Statistics}
\bvolume{50}(\bissue{5}),
\bfpage{2664}--\blpage{2693}
(\byear{2022})
\end{barticle}
\endbibitem

%%% 36
\bibitem[\protect\citeauthoryear{Lei et~al.}{2020}]{lei2020}
\begin{barticle}
\bauthor{\bsnm{Lei}, \binits{J.}},
\bauthor{\bsnm{Chen}, \binits{K.}},
\bauthor{\bsnm{Lynch}, \binits{B.}}:
\batitle{Consistent community detection in multi-layer network data}.
\bjtitle{Biometrika}
\bvolume{107}(\bissue{1}),
\bfpage{61}--\blpage{73}
(\byear{2020})
\end{barticle}
\endbibitem

%%% 37
\bibitem[\protect\citeauthoryear{Arroyo et~al.}{2021}]{arroyo2021inference}
\begin{barticle}
\bauthor{\bsnm{Arroyo}, \binits{J.}},
\bauthor{\bsnm{Athreya}, \binits{A.}},
\bauthor{\bsnm{Cape}, \binits{J.}},
\bauthor{\bsnm{Chen}, \binits{G.}},
\bauthor{\bsnm{Priebe}, \binits{C.E.}},
\bauthor{\bsnm{Vogelstein}, \binits{J.T.}}:
\batitle{Inference for multiple heterogeneous networks with a common invariant
  subspace}.
\bjtitle{The Journal of Machine Learning Research}
\bvolume{22}(\bissue{1}),
\bfpage{6303}--\blpage{6351}
(\byear{2021})
\end{barticle}
\endbibitem

%%% 38
\bibitem[\protect\citeauthoryear{Davis and Kahan}{1970}]{davis1970}
\begin{barticle}
\bauthor{\bsnm{Davis}, \binits{C.}},
\bauthor{\bsnm{Kahan}, \binits{W.M.}}:
\batitle{The rotation of eigenvectors by a perturbation. iii}.
\bjtitle{SIAM Journal on Numerical Analysis}
\bvolume{7}(\bissue{1}),
\bfpage{1}--\blpage{46}
(\byear{1970})
\end{barticle}
\endbibitem

%%% 39
\bibitem[\protect\citeauthoryear{Tong et~al.}{2024}]{tong2023uniform}
\begin{botherref}
\oauthor{\bsnm{Tong}, \binits{X.T.}},
\oauthor{\bsnm{Wang}, \binits{W.}},
\oauthor{\bsnm{Wang}, \binits{Y.}}:
Uniform error bound for {PCA} matrix denoising.
Bernoulli
(2024)
\end{botherref}
\endbibitem

%%% 40
\bibitem[\protect\citeauthoryear{Stehl{\'e} et~al.}{2011}]{stehle2011high}
\begin{barticle}
\bauthor{\bsnm{Stehl{\'e}}, \binits{J.}},
\bauthor{\bsnm{Voirin}, \binits{N.}},
\bauthor{\bsnm{Barrat}, \binits{A.}},
\bauthor{\bsnm{Cattuto}, \binits{C.}},
\bauthor{\bsnm{Isella}, \binits{L.}},
\bauthor{\bsnm{Pinton}, \binits{J.-F.}},
\bauthor{\bsnm{Quaggiotto}, \binits{M.}},
\bauthor{\bsnm{Broeck}, \binits{W.}},
\bauthor{\bsnm{R{\'e}gis}, \binits{C.}},
\bauthor{\bsnm{Lina}, \binits{B.}}, \betal:
\batitle{High-resolution measurements of face-to-face contact patterns in a
  primary school}.
\bjtitle{PloS one}
\bvolume{6}(\bissue{8}),
\bfpage{23176}
(\byear{2011})
\end{barticle}
\endbibitem

\end{thebibliography}

\end{document}